\documentclass[aps,prl,amsmath,amssymb,superscriptaddress,preprint]{revtex4-1}
\pdfoutput=1
\usepackage{graphicx}
\usepackage{grffile} %needed when image file name has multiple dots
\usepackage{dcolumn}
\usepackage{bm}
\usepackage{subfigure}
\usepackage{amssymb}
\usepackage{amsmath}
\usepackage{natbib}
\usepackage[pdftex]{hyperref}
\hypersetup{colorlinks=true,citecolor=blue,linkcolor=red,urlcolor=blue}
\usepackage[all]{hypcap}
\usepackage{color}
\definecolor{red}{rgb}{0.8, 0.0, 0.0}
\definecolor{blue}{rgb}{0.06, 0.2, 0.65}
\definecolor{green}{rgb}{0,0.6,0}

\begin{document}

\title{Automatic Design of Mechanical Metamaterial Actuators}
\author{Silvia Bonfanti}
\author{Roberto Guerra}
\author{Francesc Font Clos}
\author{Daniel Rayneau-Kirkhope}
\affiliation{Center for Complexity and Biosystems, Department of Physics, University of Milan, via Celoria 16, 20133 Milano, Italy}
\author{Stefano Zapperi}
\affiliation{Center for Complexity and Biosystems, Department of Physics, University of Milan, via Celoria 16, 20133 Milano, Italy}
\affiliation{CNR - Consiglio Nazionale delle Ricerche,  Istituto di Chimica della Materia Condensata e di Tecnologie per l'Energia, Via R. Cozzi 53, 20125 Milano, Italy}

\begin{abstract}
Mechanical metamaterials actuators achieve pre-determined input--output
operations exploiting architectural features encoded within a single
3D printed element, thus removing the need of assembling different
structural components. Despite the rapid progress in the field, there is still a need for efficient strategies to optimize metamaterial design for a variety of functions.
We present a computational method for the automatic design of mechanical metamaterial actuators
that combines a reinforced Monte Carlo method with discrete element simulations. 3D printing of selected mechanical metamaterial actuators shows that the machine-generated structures can reach high efficiency, exceeding human-designed structures. We also show that it is possible to design efficient actuators by training a deep neural network, eliminating the need for lengthy
mechanical simulations. The elementary actuators devised here can be combined to produce metamaterial machines of arbitrary complexity for countless engineering applications.
\end{abstract}

\maketitle

\section*{Introduction}

Mechanical metamaterials are a novel class of artificial materials engineered to have exceptional properties and responses that are difficult to find in conventional materials  \cite{Zheng2014}. Their stiffness, strength-to-weight ratio \cite{Farr2008}, elastic response \cite{Coulais2017} or Poisson’s ratio \cite{Lakes1987,Bertoldi2010,babaee2013}
can be tuned to match or exceed those found in standard materials. Metamaterials derive their properties not from the inherent nature of the bulk materials, but from their artificially designed internal geometry composed of multiple sub-elements, or cells, which are usually arranged in repeated regular patterns \cite{gibson2005biomechanics}. Since cells can be designed and placed in countless different ways, the resulting structure can display many degrees of freedom, giving rise to a variety of unusual physical properties which then find natural applications in industrial design, as architectural motifs or reinforcement patterns for textiles, beams and other objects. The increasing interest on metamaterials is also stimulated by the recent advances in digital manufacturing technologies e.g. 3D printing and automated assembly, which enable rapid manufacture of such material structures with the removal of many of the constraints in scale and geometry at lower and decreasing cost \cite{Berger2017}.

Metamaterials can also be considered as true machines \cite{Ion2016}, able to accomplish mechanical functions through the transformation of input stimuli into a programmable set of outputs. In this case, constituent cells work together in a well-defined manner to obtain the final controlled directional macroscopic movement. Metamaterial machines can be exploited as mechanical actuators for human-machine interactions or as interactive/responsive components
in robotics. Conventional design strategies for metamaterial structures and machines are
often based on manual operations pushed by human wills, which work reasonably well in specific conditions but are not guaranteed to yield maximum efficiency for all cases.

To overcome the limitations of passive manual design, we propose an automatic optimization method
for the automatic design of  mechanical metamaterial actuators (MMA), searching for the optimal output response to an applied input through iterative modifications of the structure.
Generative methods based on optimization algorithms have been used to design allosteric materials \cite{Rocks2017,Yan2017,Yan2018}, fiber-reinforced actuators \cite{Connolly2016} and to choose the optimal cell geometry in periodic metamaterial lattices \cite{schwerdtfeger2011design,Sharpe2018}.
Our method deals instead with ordered or disordered \cite{hanifpour2018mechanics,rayneau2019density}metamaterials and selects the optimal output response to a given applied input through iterative modifications by removal or reinsertion of beams. We show that this optimization process can be  efficiently realized coupling the optimization algorithm with discrete element simulations or with a suitably trained deep neural network.

\begin{figure*}[tbh!]
\centering
\includegraphics[width=\textwidth]{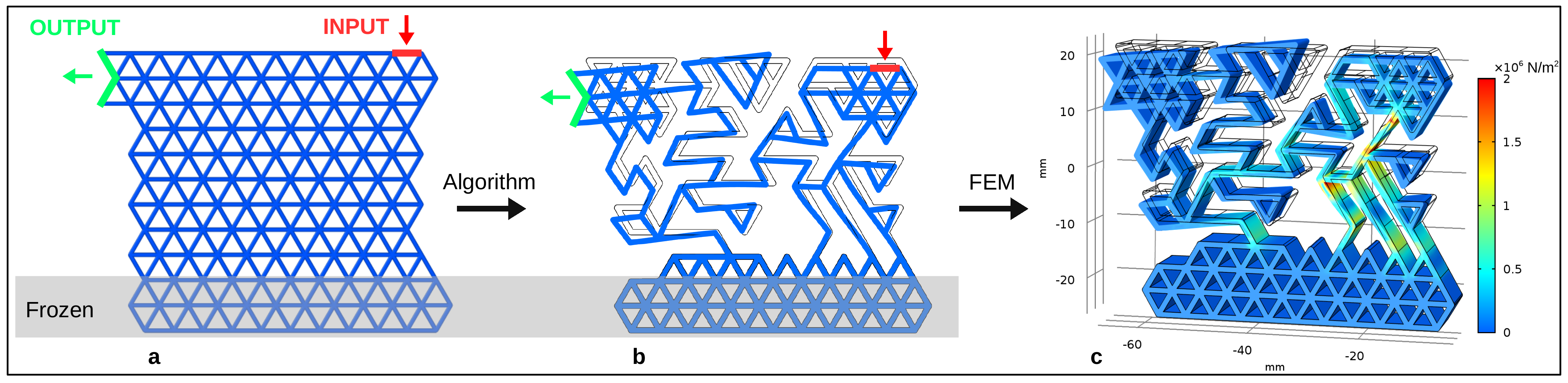}
\caption{{\bf Schematic of automatic structure generation}. a) The initial triangular lattice configuration ${\bf R}_{IS}$. b) The optimized structure obtained with DEM and its modeled response upon input displacement. c) The same structure movement simulated by FEM.
}\label{fig.model}
\end{figure*}

\section*{Methods}

\subsection*{Initial conditions}

We start from a triangular lattice configuration, which in the following we will refer to as ``Inherent Structure''~\cite{weber1984inherent} with coordinates ${\bf R}_{IS}$. Such configuration (see Figure~\ref{fig.model}a) is mechanically stable and consists of $n$ beams of length $r=r_0$ connected to $N$ nodes. The position of the $i$-th node is ${\bf r}_i=(x_i,y_i)$, $\lbrace {\bf r}_i \rbrace=({\bf r}_1,{\bf r}_2,..,{\bf r}_N)$, and the distance between two nodes is ${\bf r}_{ij}=\vert {\bf r}_j-{\bf r}_i \vert$. We then select two far-apart (group of) nodes $i$ and $j$ which represent input and output regions, respectively, and we define two normalized vectors identifying their desired direction ${\bf t}_{\mathrm{inp}}$ and ${\bf t}_{\mathrm{out}}$.

\subsection*{Efficiency}
The response of the metamaterial is monitored through its efficiency $\eta$,
\begin{equation}\label{eq1}
\eta = \frac{ {\bf t}_{\mathrm{out}}\cdot({\bf r}_j-{\bf r}_{0j}) }{ {\bf t}_{\mathrm{inp}}\cdot({\bf r}_i-{\bf r}_{0i}) }
\end{equation}
where $({\bf r}_j-{\bf r}_{0j})$ represents the displacement of the output nodes from their original positions and the dot products are averaged over the input and output number of nodes.
One can envise alternative definitions of efficiency, suitable to enforce the desired response in terms of a specific requirements of the optimized structure. We can provide two alternatives.

%\begin{itemize}
\paragraph{Direction-based efficiency.}  The search on metamaterial configurations is focused on the maximization of the output displacement toward the desired direction. To this purpose we generalize the dot product in Eq.~\ref{eq1} by a weight function $f(\gamma)=(2\cos(\gamma/2)^n-1)$, $n\ge2$, where $\gamma$ is the angle defined between the desired output direction ${\bf t}_{\mathrm{out}}$ and the measured one. The resulting efficiency is
\begin{equation}
\eta_d = \frac{ |{\bf r}_j-r_{0j}|~f(\gamma) }{ {\bf t}_{\mathrm{inp}}\cdot({\bf r}_i-r_{0i}) }~~,
\end{equation}
that for $n\gg2$ enforces the output motion along ${\bf t}_{\mathrm{out}}$, while for $n=2$ it is $f(\gamma)=\cos(\gamma)$, thus $\eta_d=\eta$.

\paragraph{Force-based efficiency.} In this implementation, it is required that the exerted force on the input nodes is efficiently propagated to the output nodes towards the target direction. This is especially advisable when the actuator is expected to integrate with other mechanical parts, forming a larger mechanism.
In this case, we apply a constant force on the input nodes, and we measure the force on the output nodes by means of monitoring springs, acting as dynamometers along ${\bf t}_{\mathrm{out}}$ direction. This corresponds to adding the energy term $E_{\mathrm{inp}}=F_{\mathrm{ext}}[{\bf t}_{\mathrm{inp}}\cdot({\bf r}_i-{\bf r}_{0i})]$ to the input nodes, and $E_{\mathrm{out}}=\frac{1}{2}k_{\mathrm{ext}}({\bf t}_{\mathrm{out}}\cdot ({\bf r}_j-{\bf r}_{0j}))^2$ to the output ones during energy minimization, with $F_{\mathrm{ext}}$ being the constant input force, and $k_{\mathrm{ext}}$ being the spring constant of the output monitor springs.
The force-based efficiency can then be straightfowardly defined as
\begin{equation}
\eta_f = \frac{ k_{\mathrm{ext}} |{\bf r}_j-{\bf r}_{0j}|f(\gamma) }{F_{\mathrm{ext}}}.
\end{equation}
%\end{itemize}

\subsection*{Optimization}
Once a suitable efficiency function is chosen, we maximize it by minimizing the cost function $\Delta=\exp(-\eta)$.
The minimization protocol makes use of the Monte Carlo (MC) method combined with conjugate gradient (CG) optimization: at each iteration step, from the present configuration with $\Delta=\Delta^{0}$, a trial configuration is obtained by removing or readding a randomly-selected beam. Input nodes, output nodes, and the three most bottom rows of nodes are discarded from pruning -- the latter are also constrained against motion (see Fig.~\ref{fig.model}).
We then displace the input nodes in the ${\bf t}_{\mathrm{inp}}$ direction (or apply an external force to them in the case of $\eta_f$), perform a CG relaxation, and by measuring the displacement of the output nodes (or the force on them through the monitoring springs) we evaluate the corresponding $\Delta^{trial}$. If $\Delta^{trial} < \Delta^{0}$ the removal/readding of the beam is accepted, otherwise it is accepted with a probability $P=exp[-(\Delta^{trial}-\Delta^{0})/T]$, where $T$ is a parameter acting as ``temperature'' in the MC dynamics.

\begin{figure}[b!]
\centering
\includegraphics[width=\columnwidth]{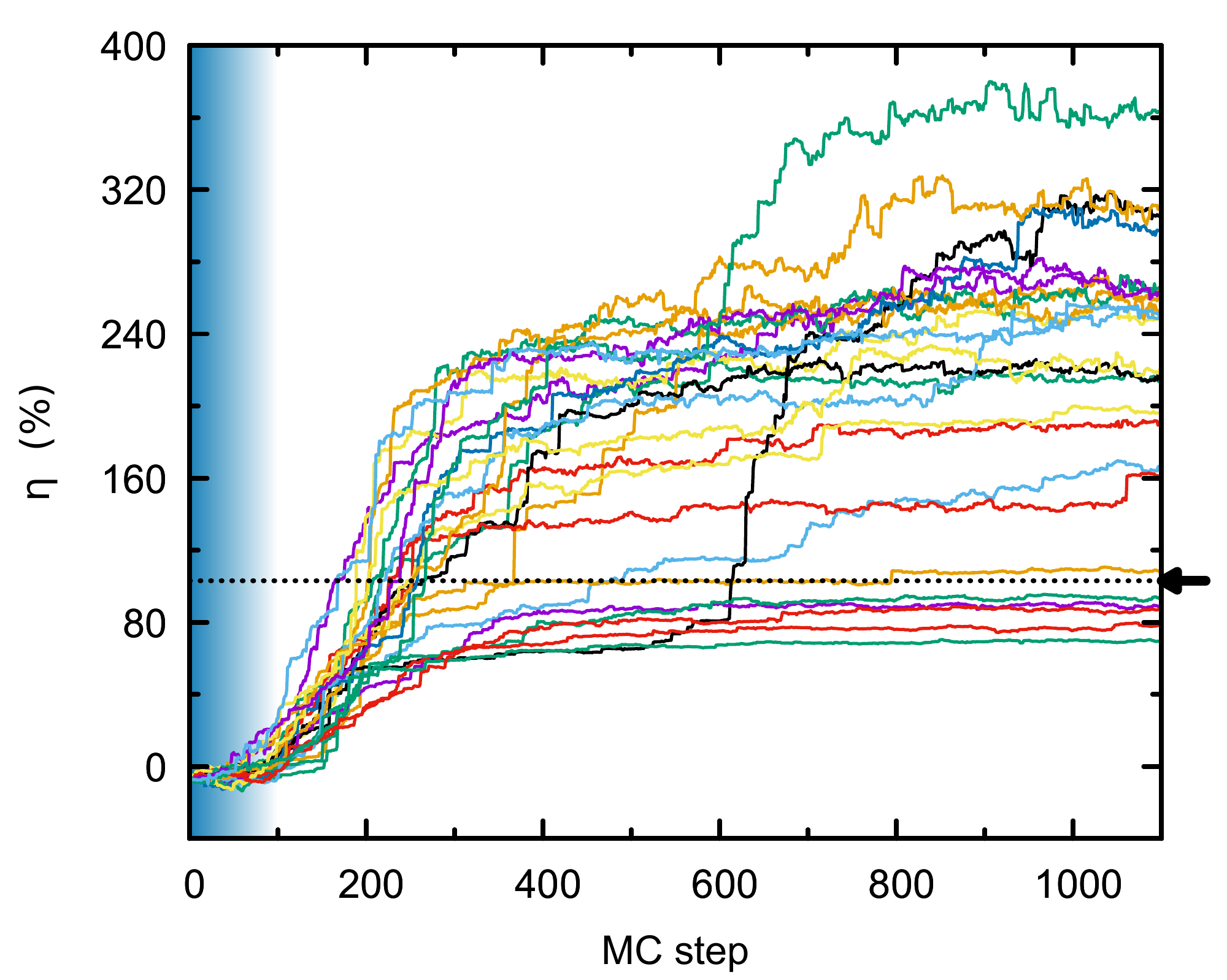}
\caption{{\bf Efficiency evolves across different paths.} Plot of instantaneous $\eta$ during MC dynamics for the orthogonal motion case. Each curve is obtained with a different initial random seed. Shaded area highlights the annealing phase in the first 100 MC steps (see Method). Only the efficiencies corresponding to accepted MC steps are reported. Arrow and dashed line mark the efficiency of the human-designed structure.
}\label{fig.eff.plot}
\end{figure}

At the beginning of each minimization, in order to explore the complex efficiency landscape \cite{Kirkpatrick1983}, we perform 100 (accepted) MC steps of annealing with $T$ linearly decreasing from $T=0.06$ (which is the threshold to consistently get $P \simeq 1$) to $T=10^{-3}$, and we subsequently let the algorithm evolve at the latter temperature. The whole procedure has been repeated in several runs using different random-number-generator seeds (see Fig.~\ref{fig.eff.plot}).

It is worth to note that in a system with $N_b$ beams, the configuration space counts $\sim2^{N_b}$ possible structures, in our system being $2^{203}\simeq10^{61}$. Clearly, such exponential scaling with $N_b$ severely limits the possibility of full exploration of the configuration space, and thus very fast methods to predict the efficiency of the trial structures are required. To maximize such exploration we have employed a combination of three different methods acting at different approximation levels, as described in the following.

\subsection*{I - Discrete Element Model}
\begin{figure}[t!]
\centering
\includegraphics[width=\columnwidth]{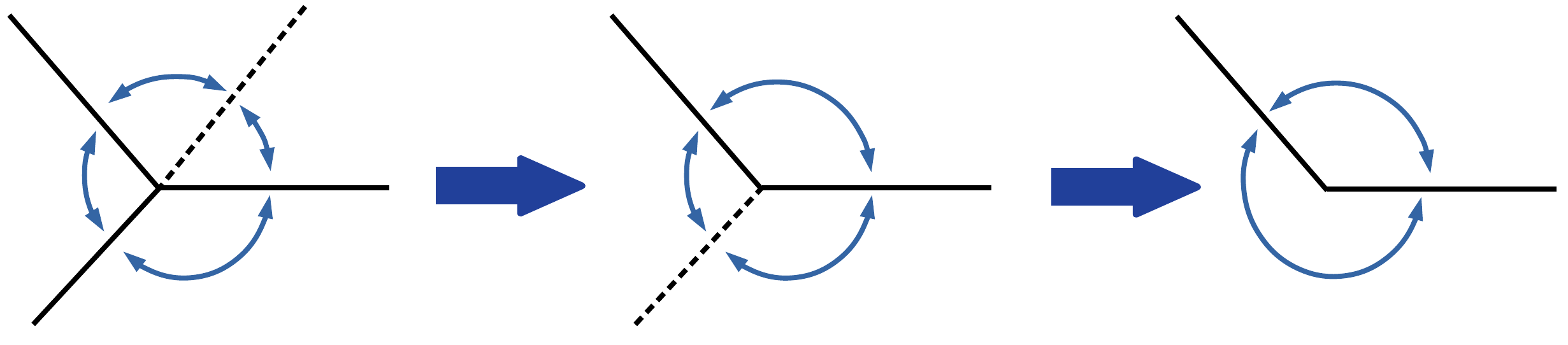}
\caption{Angular springs re-neighboring. When a beam is removed (dashed line), the two angular springs connecting to that beam are removed, and a new spring is formed between the newly neighboring beams.
}\label{fig.sketch_angle}
\end{figure}

To obtain a fast and reliable estimation of the efficiency of a given structure we have made use of a simplified discrete element model (DEM) of the lattice, in which the total energy can be expressed as
\begin{equation}
E=\sum_i \sum_{j>i} \phi_2({\bf r}_{ij})+\sum_i \sum_{j>i} \sum_{k\neq j} \phi_3({\bf r}_{ij},{\bf r}_{ik},\theta_{ijk})
\end{equation}
where the pairwise term is a spring potential with rest length $r_0$,
\begin{equation}
\phi_2({\bf r}_{ij})=k ( {\bf r}_{ij}-{\bf r}_0 )^2~~,
\end{equation}
while the 3-body term introduces angular springs among the nearest-neighbor beams connected to the same node,
\begin{equation}
\phi_3({\bf r}_{ij},{\bf r}_{ik},\theta_{ijk})=\lambda \big[\theta_{ijk} - \theta^0_{ijk}\big]^2
\label{3body_sw}
\end{equation}
being $\theta_{ijk}$ the angle formed by beams $\overline{ij}$ and $\overline{ik}$, and $\theta^0_{ijk}$ the initial angle value in the triangular lattice.
Both $\phi_2$ and $\phi_3$ act among first neighbors only, with 3-body neighbors dynamically recalculated at each step (see Fig.~\ref{fig.sketch_angle}). If not stated differently, we have employed the following unit-less parameters: $k=5$, $\lambda=0.1$, $r_0=1$.

The minimization of $E$ in the presence of frozen nodes (acting as basement) and of displaced input nodes, allows us to predict the response of the trial structure with a good compromise between speed and reliability. This will be therefore our reference method for the seek of structures with the highest efficiency (see Figure~\ref{fig.model}b).

\subsection*{II - Finite Element Method}

A realistic simulation of the mechanical response of a structure subject to an external force can be obtained by the finite element method (FEM), which describes the macroscopic response of the system as sum of the responses of a large number of its sub-elements. The method is very accurate in the elastic regime, but works on the timescale of several seconds for our reference system (and greatly growing with the system size), and it is thus not suitable for the trial-error MC search of efficient structures. Rather, FEM has been employed before 3D printing as a final validation of the selected structures, generated through the more simplified and fast methods described above.

3D models of simulated structures have been produced by extrusion of each bond (see Figure~\ref{fig.model}c). For FEM calculations we have employed {\small COMSOL} Multiphysics and {\small COMSOL with MATLAB} through its structural mechanics module \cite{comsol}. All studies assume a linear elastic material with Young’s modulus and Poisson’s ratio estimated experimentally for bulk samples of interest. Results are obtained using Euler-Bernoulli beam elements and the in-built ``stationary studies'' calculation (a quasi-static solver). 

\subsection*{III - Machine Learning}
\textit{Deep Learning Model.} We take the Resnet50 architecture as implemented in the python Keras library \cite{Chollet2015-hm} and repurpose it to perform regression instead of classification by modifying the top layer with a single-unit dense layer. \\
\textit{Data preparation.} We generate images of configurations from two thousand metropolis runs, half of which aim at positive efficiency and half of which aim at negative efficiency, including always both accepted and rejected configurations. In total, we generate 192$\times$128 pixel PNG images of $1163733$ configurations using the Python plotting library matplotlib~\cite{Hunter2007-cl}.\\
\textit{Model training.} We use a standard Adam optimizer and a mean squared error loss function.
75\% of the runs are used as training data, while the remaining 25\% are used for validation. We also implement a simple resampling strategy that renders the distribution of efficiencies approximately uniform over the training data. This mitigates the fact that, otherwise, the model predictions would be less accurate for very high-efficiency configurations, since these are less commonly found in our dataset. We train our modified Resnet50 CNN for 10 epochs in batches of size 32.\\
\textit{Computational environment and model performance} We use a 16-core computer equiped with a Tesla K20c GPU for all CNN computations.
We measure the performance of the model using the $R^{2}$ value of linear regression between actual and predicted efficiency values. %The model achives a remarkable value of $R^2 = 0.966$ on the test set.

\subsection*{Human designed structures and 3D printing}

To provide a comparison of the automatically designed MMA with conventional solutions, human-designed counterparts of the MMAs have been created before running any simulation, as to avoid any possible bias on the designer about the mechanisms leading to a high efficiency.

Samples are then produced by means of 3D printing, using the fused deposition modeling technique. In this method, the final structure is produced by laying down many successive thin layers of molten plastic. Each layer thickness is $0.8 n_s$, where the nozzle size is $n_s=0.4mm$. The material used is NinjaFlex, a formulated Thermoplastic Polyurethane (TPU) material with super elastic properties.

\section*{Results}
\subsection*{Automatic design achieves high efficiency}

\begin{figure}[b!]
\centering
\includegraphics[width=\columnwidth,angle=0,page=2]{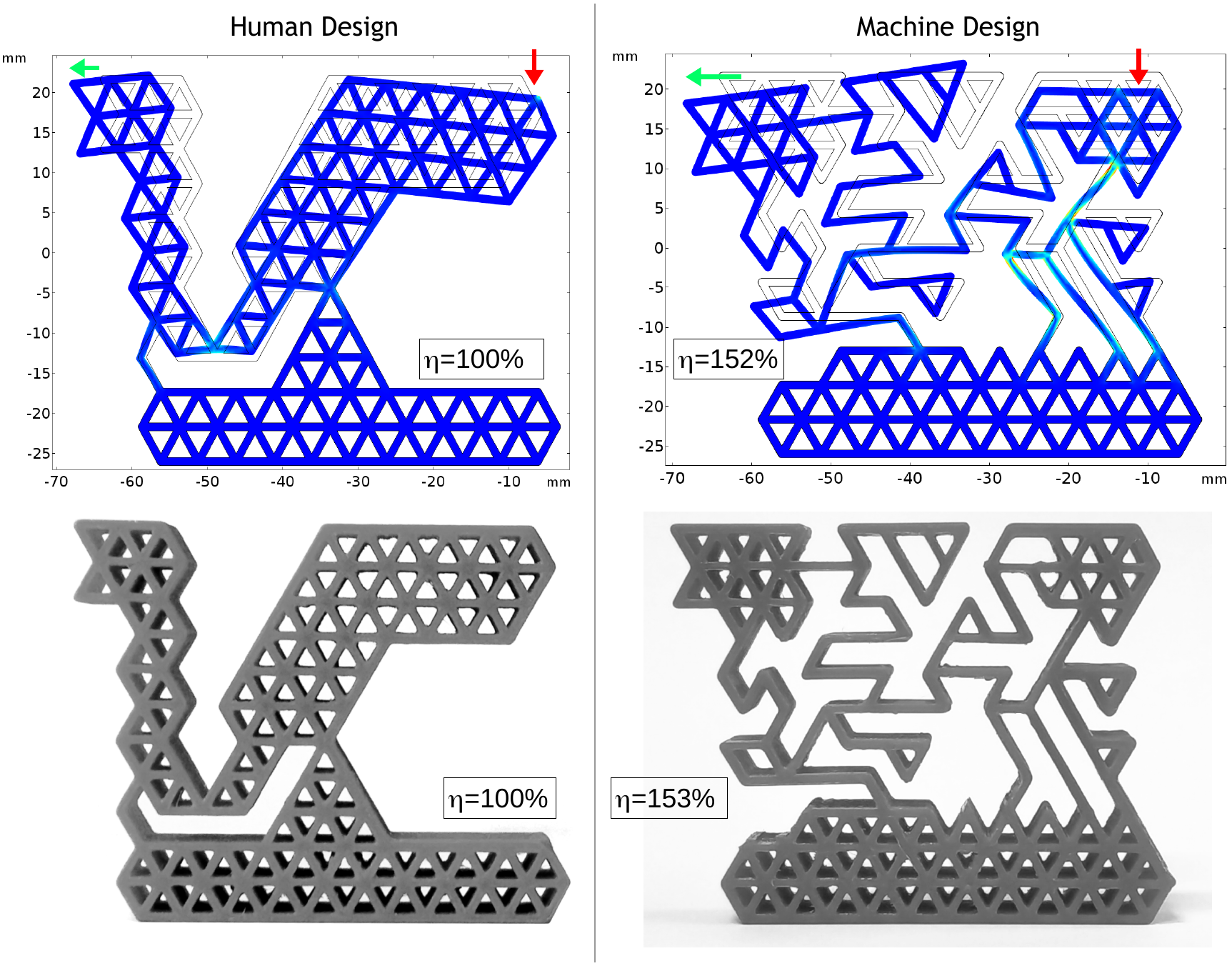}
\caption{{\bf Automatic design achieves efficient anti-parallel orthogonal movement}. Comparison of (left panels) human-designed and (right panels) machine-designed structures for (top panels) FEM simulated and (bottom panels) the corresponding 3D printed realizations. Color gradient reports the stress with the same colorbar as in Fig.~\ref{fig.model}c. Resulting efficiencies are reported in each panel. For movement visualization see Supplementary Movie 2.
}\label{fig.crane}
\end{figure}

We consider two prototypical actuators, the first in which the desired input and output are orthogonal, ${\bf t}_{\mathrm{inp}}= -\hat{y}$, ${\bf t}_{\mathrm{out}}= -\hat{x}$, and a second one in which they are anti-parallel, ${\bf t}_{\mathrm{inp}} = -\hat{y}$, ${\bf t}_{\mathrm{out}}= \hat{y}$. In Figure~\ref{fig.eff.plot} we report the evolution of $\eta$ during the MC dynamics for several realizations of the orthogonal-functionalized MMAs (traces for the anti-parallel case are similar). An example of generation of such structure during the optimization steps can be visualized in Supplementary Movie 1. We note that, after the annealing phase, efficiency tends to evolve in steps, interspersed by noisy parts and by plateaus. Steps can occur when a mechanism that engages the desired response is finally triggered, while plateaus indicate a structure whose efficiency is robust against the removal or addition of one or more bonds. Hence, we have selected our most efficient reference samples from the configurations laying in these plateaus of $\eta$.
We obtain a large variability of the final $\eta$ values, indicating a possible trapping in local minima that require ``thermally activated'' excitation to escape. Because of this, and depending on the post-annealing conditions, the exploration of the whole phase-space of the network can take very long times, which further increase with the network size.

\begin{figure}[t!]
\centering
\includegraphics[width=\columnwidth,angle=0,page=1]{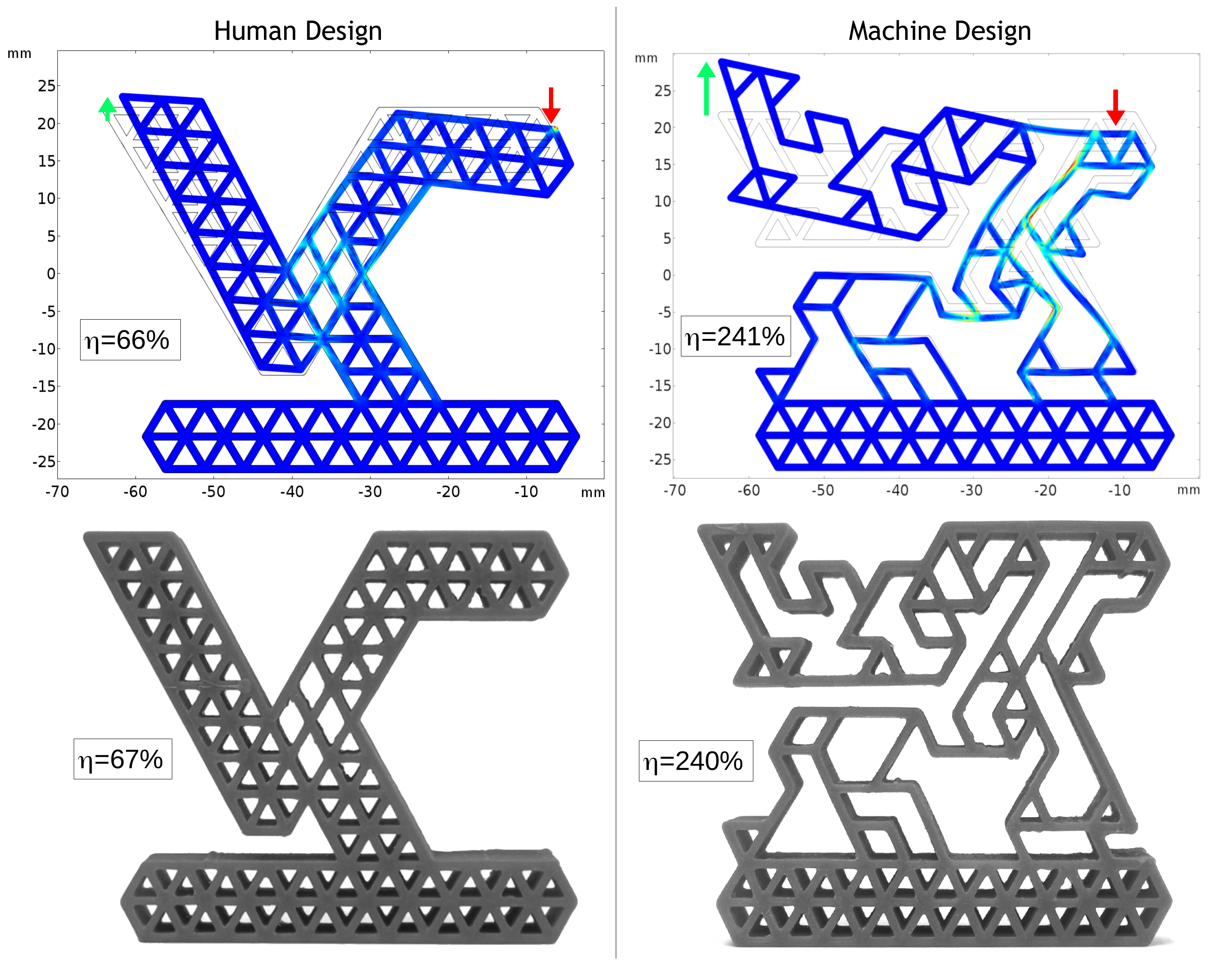}
\caption{{\bf Automatic design achieves efficient anti-parallel movement.} Comparison of (left panels) human-designed and (right panels) machine-designed structures for (top panels) FEM simulated and (bottom panels) the corresponding 3D printed realizations. Color gradient reports the stress with the same colorbar as in Fig.~\ref{fig.model}c. Resulting efficiencies are reported in each panel. For movement visualization see Supplementary Movie 3.
}\label{fig.button}
\end{figure}

The use of a simplified model within the minimization algorithm requires further validation through more refined simulations. We thus have converted our structures to a FEM mesh, so to simulate the realistic response of the material (see Method). Simulation results for orthogonal and anti-parallel motion are reported in Figures~\ref{fig.crane} and \ref{fig.button}, respectively.
Note that the achievable efficiency can easily approach and exceed the value obtained by the human-designed structures, reported on the left panels.
In all the cases considered, the model-calculated efficiencies were validated by FEM simulations, with minimal corrections on $\eta$ with respect of DEM estimates. A further advantage of FEM simulations is the possibility to obtain information about the stress propagation along the network, and to identify the regions mostly involved in the mechanism actualization. In this respect, we note that the machine-generated structures are characterized by a broad distribution of the stress, indicating a collective engagement of the network. Conversely, in the human-designed structures the stress highlights the few pivot points employed to perform the movement.

\subsection*{Efficiency is predicted by machine learning}
The large set of metamaterial configurations we obtain offer a possibility for further insight when regarded as a dataset to be inquired. In this setting, one can naturally pose several questions related to how changes in structure relate to changes in efficiency.
Here we assess whether static images of the configurations can be used to infer their efficiencies without the need of performing DEM or FEM simulations. To respond to this question, we have trained a Convolutional Neural Network (CNN) to perform image regression, that is, to predict the efficiency of a configuration from an image of its layout. Notice how this differs from the more standard use of CNN for image classification. Using a large number of configurations ($N\simeq 10^6$), we have been able achieve an accuracy of $R^2 = 0.966$, see Fig.~\ref{fig.ML}(a,c) and Methods for details.

\subsection*{New structures can be generated through machine learning}
\begin{figure*}[tbh]
\centering
\includegraphics[width=\textwidth]{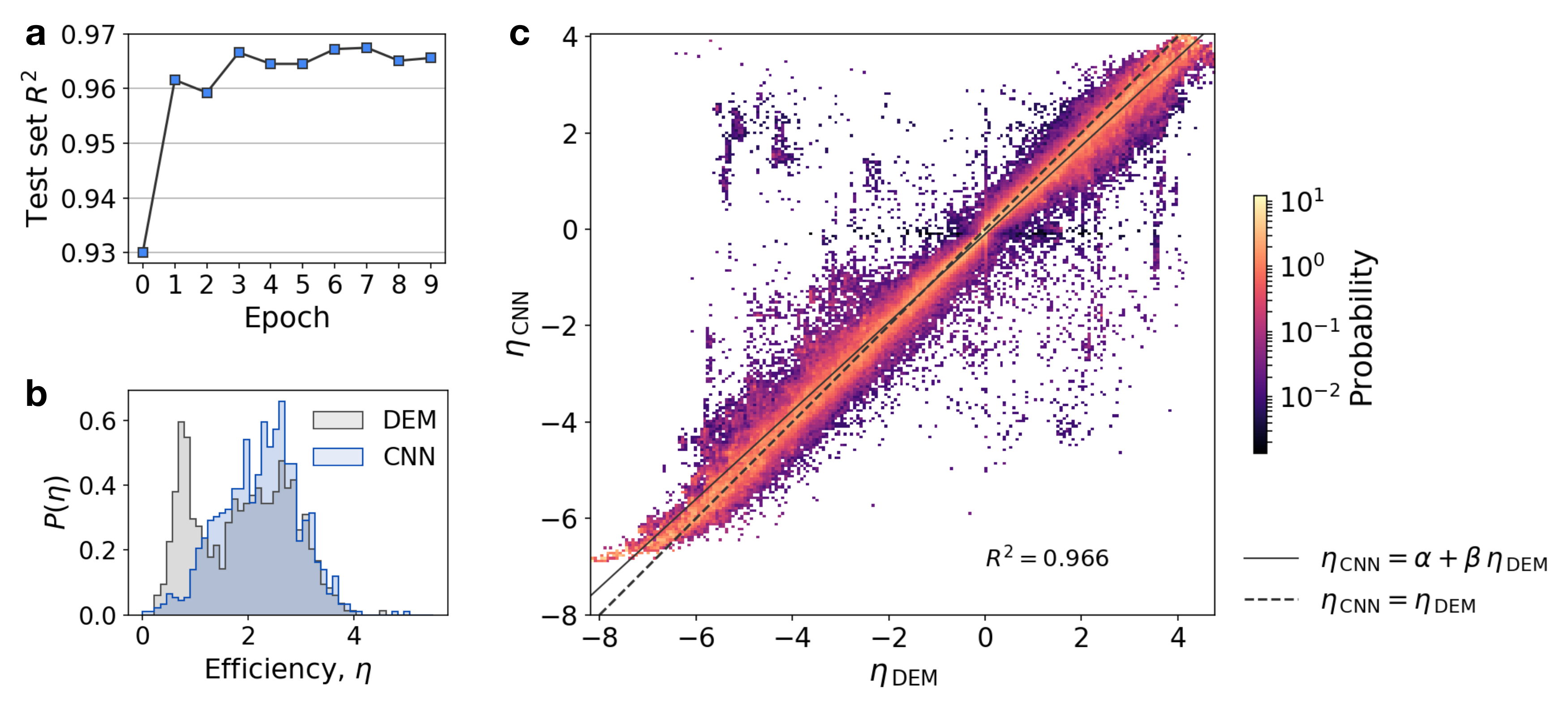}
\caption{{\bf Machine learning can be used to predict efficient and design structures}
(a) Coefficient of determination $R^2$ of the test set during 10 training epochs. (b) Distribution of efficiency for configurations generated with the DEM model (gray) and the CNN model (blue). (c) Density plot of the efficiency $\eta_\mathrm{DEM}$ (measured by DEM model)  and $\eta_\mathrm{CNN}$ (predicted by CNN model) showing that the CNN model is able to predict the DEM model efficiency values well. The solid line represent a linear fit, whose coefficient of determination is $R^2=0.966$. The dashed line is included as a guide to the eye only.
}\label{fig.ML}
\end{figure*}

Finally, we take our ML-framework one step further and ask whether we can use our CNN model to generate new configurations from scratch by substituting the spring-mass model efficiency estimation step with predictions from the CNN model. In other words, we use the same Monte Carlo strategy described above but, at each step, instead of measuring the efficiency of the proposed configuration with spring-mass model or finite-element method, we estimate it with the CNN model. Interestingly, the procedure is successful and we are able to generate new, efficient configurations with an approximately 100-fold speed-up with respect to spring-mass model. In addition, the CNN-generated configurations have a distribution of efficiency similar to that of the metropolis-generated ones, see Fig. \ref{fig.ML}(b). To ensure that the obtained configurations are different from the ones used to train the CNN, we measure the minimal distance between each ML-generated configuration and all the configurations used during training, and find that the typical distance is of around 65 bonds. We regard these results as an exploratory, first step towards ML-aided generation of metamaterials.

\section*{Discussion}
In this paper, we have proposed algorithms for the automatic design of MMAs with a broad set of possible movements and efficiency – exceeding those of human-designed solutions. In its first implementation, the algorithm exploits a discrete elements model to obtain an approximation of the mechanical efficiency of the structures which is then
used to drive a Monte Carlo search over the possible structures. FEM calculations are then used
to confirm the efficiency of the optimized structures that can then be realized by
3D printing. Here we have concentrated on two dimensional actuators but an extension to fully three dimensional models is also possible using the same strategy employed here.

Furthermore, we have employed deep neural networks to predict the efficiency of the
actuator from its structure. Once properly trained, the neural network can be used to
create new structures without the need of performing DEM or FEM simulations.
The use of machine learning to assist the automatic design of MMAs opens intriguing
possibilities in terms of algorithmic speed, because it could
potentially allow to design larger structures that can not be efficiently simulated
by DEM.

In conclusions, our work constitute the first step toward the establishment of a reference library of elementary actuators (EA). More complex actuators can be subsequently obtained by the interlinking of multiple EA, with countless possibilities in terms of applications and flexibility. For instance, the algorithms could be useful to design moving
parts in machines and robots, especially at small scales where the surface-to-volume ratio is very large, thus leading to dominant friction and wear. Benefits spans from the availability of ready-to-use EA, that will constitute a reference to engineers and material scientists, to the possibility of providing custom solutions for non-standard applications.

\section*{Acknowledgements}
This research has been supported by the European Research Council through the
Proof of Concept grant 841640 METADESIGN. We gratefully thank F.\ Pezzotta for assistance with 3D printing. 

%\small
%\bibliographystyle{unsrt}
%\bibliography{biblio}

\begin{thebibliography}{10}

\bibitem{Zheng2014}
X.~Zheng, H.~Lee, T.~H. Weisgraber, M.~Shusteff, J.~DeOtte, E.~B. Duoss, J.~D.
  Kuntz, M.~M. Biener, Q.~Ge, J.~A. Jackson, S.~O. Kucheyev, N.~X. Fang, and
  C.~M. Spadaccini.
\newblock Ultralight, ultrastiff mechanical metamaterials.
\newblock {\em Science}, 344(6190):1373--1377, June 2014.

\bibitem{Farr2008}
RS~Farr and Yong Mao.
\newblock Fractal space frames and metamaterials for high mechanical
  efficiency.
\newblock {\em EPL (Europhysics Letters)}, 84(1):14001, 2008.

\bibitem{Coulais2017}
Corentin Coulais, Chris Kettenis, and Martin van Hecke.
\newblock A characteristic length scale causes anomalous size effects and
  boundary programmability in mechanical metamaterials.
\newblock {\em Nature Physics}, 14(1):40--44, September 2017.

\bibitem{Lakes1987}
Roderic Lakes.
\newblock Foam structures with a negative poisson's ratio.
\newblock {\em Science}, pages 1038--1040, 1987.

\bibitem{Bertoldi2010}
Katia Bertoldi, Pedro~M Reis, Stephen Willshaw, and Tom Mullin.
\newblock Negative poisson's ratio behavior induced by an elastic instability.
\newblock {\em Advanced materials}, 22(3):361--366, 2010.

\bibitem{babaee2013}
Sahab Babaee, Jongmin Shim, James~C Weaver, Elizabeth~R Chen, Nikita Patel, and
  Katia Bertoldi.
\newblock 3d soft metamaterials with negative poisson's ratio.
\newblock {\em Advanced Materials}, 25(36):5044--5049, 2013.

\bibitem{gibson2005biomechanics}
Lorna~J Gibson.
\newblock Biomechanics of cellular solids.
\newblock {\em Journal of biomechanics}, 38(3):377--399, 2005.

\bibitem{Berger2017}
J.~B. Berger, H.~N.~G. Wadley, and R.~M. McMeeking.
\newblock Mechanical metamaterials at the theoretical limit of isotropic
  elastic stiffness.
\newblock {\em Nature}, 543(7646):533--537, February 2017.

\bibitem{Ion2016}
Alexandra Ion, Johannes Frohnhofen, Ludwig Wall, Robert Kovacs, Mirela Alistar,
  Jack Lindsay, Pedro Lopes, Hsiang-Ting Chen, and Patrick Baudisch.
\newblock Metamaterial mechanisms.
\newblock In {\em Proceedings of the 29th Annual Symposium on User Interface
  Software and Technology}, pages 529--539. ACM, 2016.

\bibitem{Rocks2017}
Jason~W. Rocks, Nidhi Pashine, Irmgard Bischofberger, Carl~P. Goodrich,
  Andrea~J. Liu, and Sidney~R. Nagel.
\newblock Designing allostery-inspired response in mechanical networks.
\newblock {\em Proceedings of the National Academy of Sciences},
  114(10):2520--2525, February 2017.

\bibitem{Yan2017}
Le~Yan, Riccardo Ravasio, Carolina Brito, and Matthieu Wyart.
\newblock Architecture and coevolution of allosteric materials.
\newblock {\em Proceedings of the National Academy of Sciences of the United
  States of America}, 114(10):2526---2531, March 2017.

\bibitem{Yan2018}
Le~Yan, Riccardo Ravasio, Carolina Brito, and Matthieu Wyart.
\newblock Principles for optimal cooperativity in allosteric materials.
\newblock {\em Biophysical Journal}, 114(12):2787--2798, June 2018.

\bibitem{Connolly2016}
Fionnuala Connolly, Conor~J. Walsh, and Katia Bertoldi.
\newblock Automatic design of fiber-reinforced soft actuators for trajectory
  matching.
\newblock {\em Proceedings of the National Academy of Sciences}, 114(1):51--56,
  December 2016.

\bibitem{schwerdtfeger2011design}
J~Schwerdtfeger, F~Wein, G~Leugering, RF~Singer, C~K{\"o}rner, M~Stingl, and
  F~Schury.
\newblock Design of auxetic structures via mathematical optimization.
\newblock {\em Advanced materials}, 23(22-23):2650--2654, 2011.

\bibitem{Sharpe2018}
Conner Sharpe, Carolyn~Conner Seepersad, Seth Watts, and Dan Tortorelli.
\newblock Design of mechanical metamaterials via constrained bayesian
  optimization.
\newblock In {\em Volume 2A: 44th Design Automation Conference}. American
  Society of Mechanical Engineers, August 2018.

\bibitem{hanifpour2018mechanics}
Maryam Hanifpour, Charlotte~F Petersen, Mikko~J Alava, and Stefano Zapperi.
\newblock Mechanics of disordered auxetic metamaterials.
\newblock {\em The European Physical Journal B}, 91(11):271, 2018.

\bibitem{rayneau2019density}
Daniel Rayneau-Kirkhope, Silvia Bonfanti, and Stefano Zapperi.
\newblock Density scaling in the mechanics of a disordered mechanical
  meta-material.
\newblock {\em Applied Physics Letters}, 114(11):111902, 2019.

\bibitem{weber1984inherent}
Thomas~A Weber and Frank~H Stillinger.
\newblock Inherent structures and distribution functions for liquids that
  freeze into bcc crystals.
\newblock {\em The Journal of chemical physics}, 81(11):5089--5094, 1984.

\bibitem{Kirkpatrick1983}
S.~Kirkpatrick, C.~D. Gelatt, and M.~P. Vecchi.
\newblock Optimization by simulated annealing.
\newblock {\em Science}, 220(4598):671--680, 1983.

\bibitem{comsol}
COMSOL Multiphysics.
\newblock Comsol multiphysics user guide (version 4.3 a).
\newblock {\em COMSOL, AB}, pages 39--40, 2012.

\bibitem{Chollet2015-hm}
Fran{\c c}ois Chollet and {Others}.
\newblock Keras.
\newblock \url{https://keras.io}, 2015.

\bibitem{Hunter2007-cl}
J~D Hunter.
\newblock Matplotlib: A {2D} graphics environment.
\newblock {\em Computing in Science Engineering}, 9(3):90--95, May 2007.

\end{thebibliography}

\end{document}